\documentclass{ws-p10x7}
\usepackage{amsmath}
\usepackage{dcolumn}
\usepackage{physics}

\newcommand{\GG}{\gamma^*\gamma^*}
\newcommand{\GGG}{\gamma\gamma}

\def\EE{\ifmath{\mathrm{e^+ e^-}}}

\begin{document}

\title{Double Tag Events in Two-Photon Collisions at LEP}

\author{M. Wadhwa}

\address{University of Basel, Klingelbergstrasse 82,\\
         CH-4056 Basel, Switzerland\\E-mail: Maneesh.Wadhwa@cern.ch}

\twocolumn[\maketitle\abstract{

  Double tag events in two photon collisions are studied using the L3 detector at the LEP
  center of mass energies $\sqrt{s} \simeq ~189-202$~GeV. The cross-section of
  $\gamma ^* \gamma ^*$ collisions is measured at an average photon virtuality
  $ \langle Q^2 \rangle = 15~\rm{GeV}^2$. The results are in agreement with Monte Carlo
  predictions based on perturbative QCD, while the Quark Parton Model alone is insufficient
  to describe the data. The measurements are compared to the LO and the NLO BFKL calculations.

}]

\section{Introduction}

  In this paper we present new results on double-tag  two-photon events
  $\EE \rightarrow \EE hadrons$. The data,
  collected at centre-of-mass energies $\sqrt{s} \simeq 189-202 \GeV$, correspond to
  an integrated luminosity of 401~pb$^{-1}$. Both scattered electrons
  \footnote{Electron stands for electron or positron throughout this paper.} are
  detected in the small angle electromagnetic calorimeters. The virtuality of the
  two photons, $Q^2_1$ and $Q^2_2$, is in the range of $4 \GeV^2 < Q^2_{1,2} < 40 \GeV^2$.

  \begin{figure}[h]
  \centering
  \includegraphics[height=0.15\textheight]{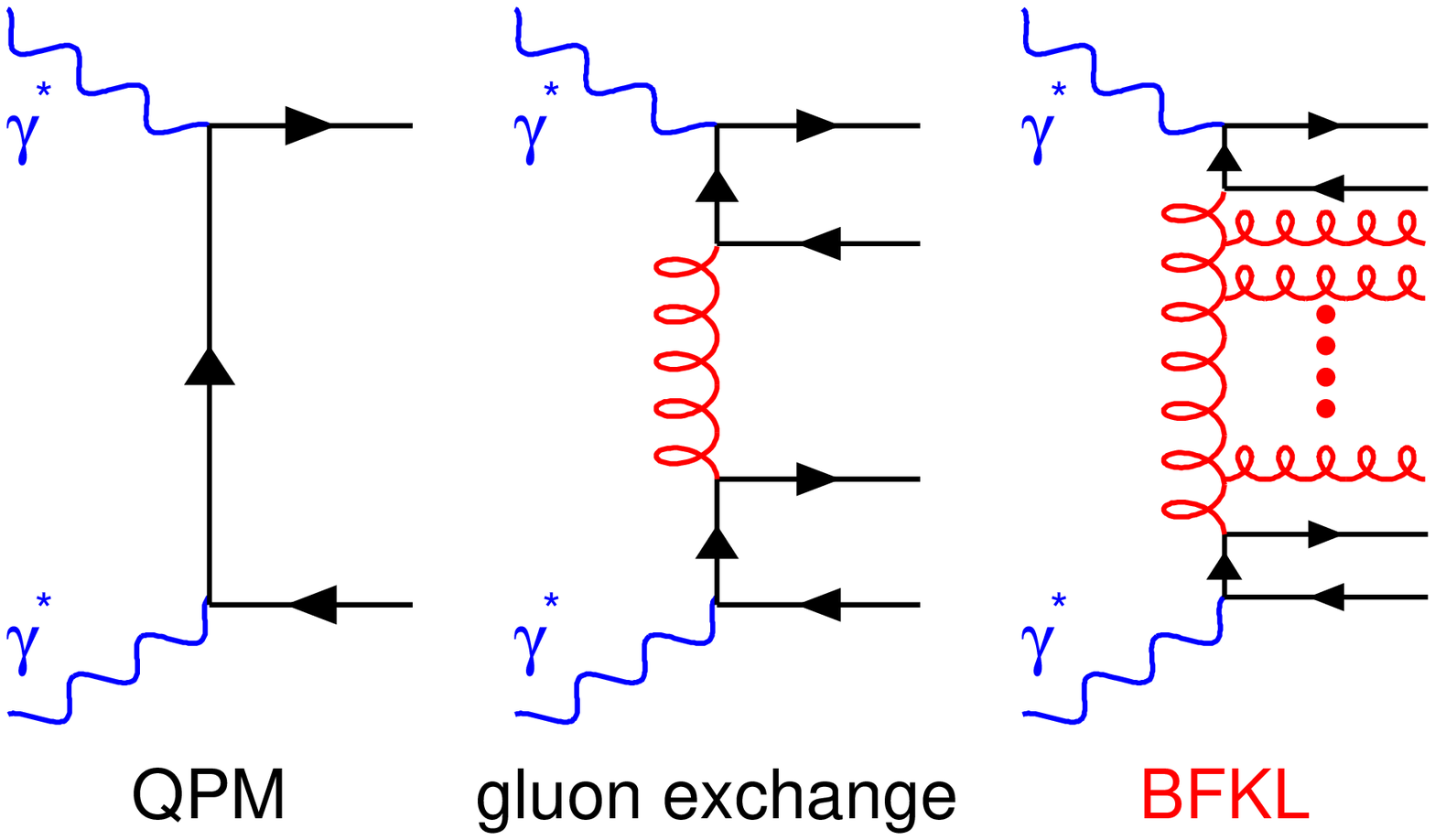}
  \caption[]{Diagrams for the a)~QPM, b)~one-gluon exchange and c)~BFKL Pomeron processes 
   in a $\GG$ interaction.
  \label{fig:feyn} }
  \end{figure}

  The centre-of-mass energy of the two virtual photons, $ \sqrt{\hat{s}}= W_{\GGG}$,
  ranges from $5 \GeV$ to $90 \GeV$. 
  The cross-section measurement of the two virtual photons is considered as "golden"
  process to test the BFKL dynamics~\cite{bfkl}. For this scheme
  the $\gamma ^* \gamma^*$ interaction can be seen as the interaction of two $q \bar{q}$
  pairs scattering off each other via multiple gluon exchange.
 ~(Fig.~\ref{fig:feyn}c). In the leading order approximation (LO), the  cross-section in the 
  saddle point approximation for the collision of  two virtual photons is~\cite{gg2,gg1}:
  \begin{equation}
    \sigma_{\GG} = \frac{\sigma_0}{{Q_1 Q_2 Y}} \left(\frac{s}{s_0}\right)^{\alpha_P-1}                   
  \end{equation}
  
  Here
  \begin{equation}
   \begin{split}
    \sigma_0 &= \text{const} \\
    s_0      &= \frac{K Q_1 Q_2 }{y_1y_2} ~~,~~  Y=\ln{(s/s_0)} \\
    y_i      &= 1-(E_i/E_{b})\cos^2(\theta_i/2)
   \end{split}
  \end{equation}
  where $E_{b}$ is the beam energy, $E_i$ and $\theta_i$ are the energy and polar angle
  of the scattered electrons and $\alpha_P$ is the ``hard Pomeron'' intercept; K is a
  scale factor which accounts for uncertainity in the BFKL energy scale $s_0$.
  The centre-of-mass energy of the two-photon system is related to the $\EE$
  centre-of-mass energy $s$ by $\hat{s} = W^2_{\GGG} \approx s y_1y_2$. In leading order
  $(\alpha_P-1)  = (4\ln2) N_c\alpha_s /\pi$, where  $N_c$ is the number of colours.
  Using $N_c = 3$ and $\alpha_s =0.2$, $(\alpha_P -1) \simeq 0.53$. The born cross-section of 
  one gluon exchange (see Fig.~\ref{fig:feyn}b) is independent of $W_{\GGG}$. Recently, effort 
  has been devoted to improve the exact leading order calculation~\cite{bfkl} by studying the 
  effect of charm mass and the contribution of longitudinal photon polarization
  states~\cite{lo}. Still these effects are not sufficient to describe our previous 
  measurement~\cite{paper_168}. One needs next to leading order corrections(NLO). It turns out 
  that the NLO corrections~\cite{nlo} to the intercept "$\alpha_P-1$" are negative for
  $\alpha_{s} > 0.16$. Different techniques~\cite{nlo1,nlo2,nlo3,nlo4,nlo5,nlo6,nlo7} have been 
  proposed to improve the NLO calculations in a suitable renormalization scheme thus giving 
  values of $(\alpha_P-1)$ in the range 0.17$-$0.33.

\section{Double-tag cross-section}

After selection cuts described in ref~\cite{note_2568}, we have selected 336 candidate events. 
The estimated background is 56 events, mainly due to  $\epem \rightarrow \epem \tau^+\tau^-$ and 
misidentified single-tag events. The contamination from annihilation processes and lepton channels 
in  two photon collisions is negligible. The preliminary cross-section is measured in the
kinematic region limited by:
 \begin{itemize}
  \item $E_{1,2} > 30 \GeV$, $30~\mathrm{mrad} < \theta_{tag} < 66$~mrad
        and $2 \leq Y \leq 7$
 \end{itemize}

\begin{figure}[htp]
  \includegraphics[height=0.26\textheight]{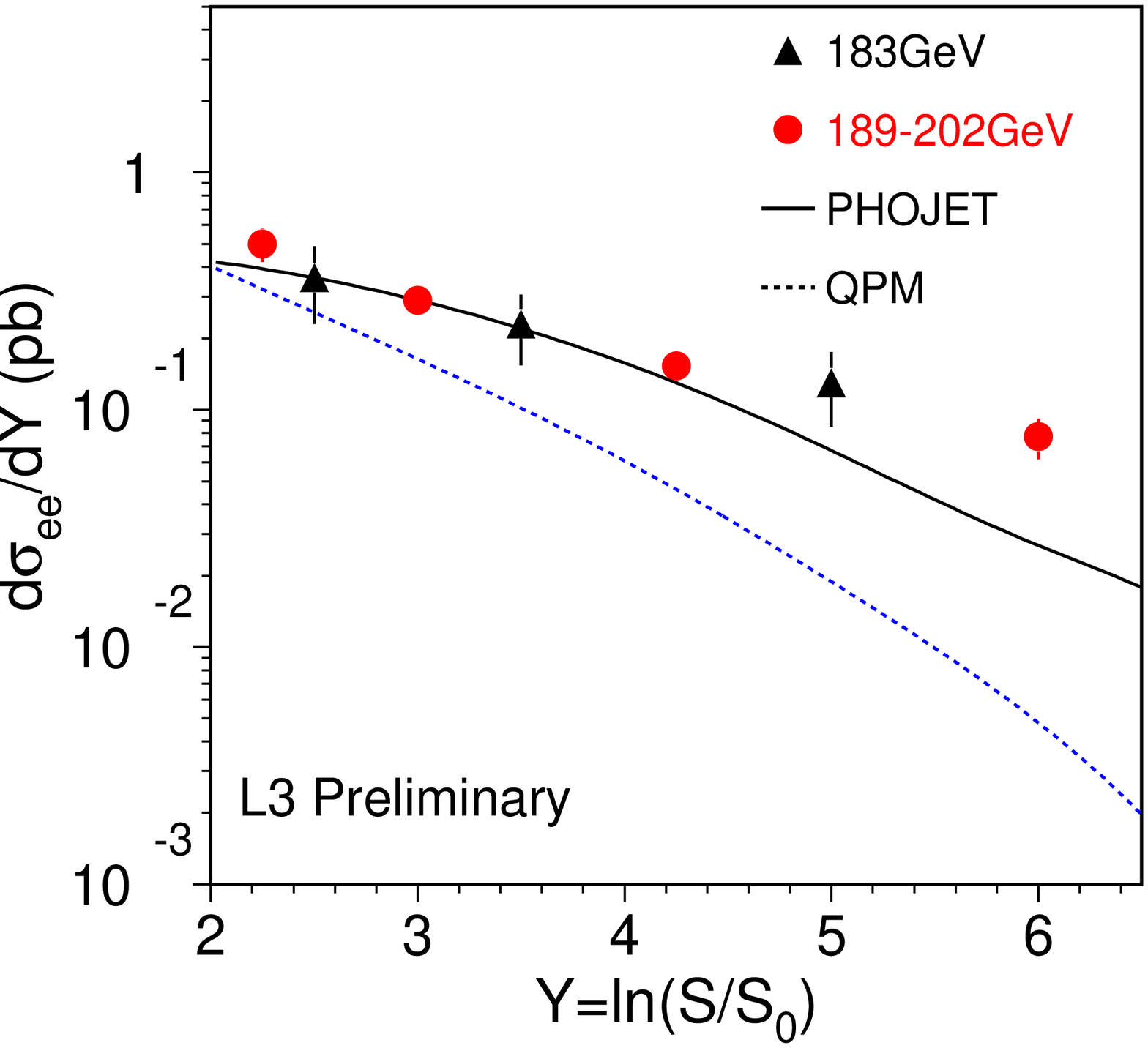}
  \caption[]{ The cross-section of $\epem \rightarrow \epem hadrons$ as a function of $Y$ in the
             kinematical region defined in the text at $\sqrt{s}\simeq 189-202\GeV$ compared to our
             previous results $\sqrt{s}\simeq 183 \GeV$. In the figure the predictions of PHOJET
             (continuos line) and of the QPM (dashed line) are indicated.
  \label{fig:cro} }
\end{figure}

\noindent
The data is then corrected for efficiency and acceptance  with two Monte-Carlo models;
PHOJET~\cite{pho} and Vermaseren(QPM)~\cite{verm} respectively. The differential cross-sections
$\mathrm{d}\sigma (\epem \rightarrow \epem + \mathrm{hadrons}) /\mathrm{d}Y$ are measured in
four $\Delta Y$ intervals. As one can be seen in Table~\ref{tab:cro} and in Fig.~\ref{fig:cro}, 
none of the models are sufficient to describe the data. The value of the  cross-section at 
5$< Y < 7$ exceeds the Monte Carlo prediction by about 3.5 standard deviations. 

\begin{table}[htp]
     \caption[]{The differential cross-section, d$\sigma (\epem \rightarrow \epem + \mathrm{hadrons})$/d$Y$ 
     in picobarn measured in the kinematic region defined in the text, at $\sqrt{s} \simeq 189 -202 \GeV$.
             The predictions of the PHOJET and the QPM Monte Carlo models are also listed. The first error is 
	     statistical and the second is systematic.
  \label{tab:cro} }
  {\setlength{\tabcolsep}{0.3mm}
  \begin{tabular}{|c|c|c|c|c|}
          \hline
                    & DATA                      & PHOJET                 &    QPM \\
         $\Delta Y$& $\mathrm{d\sigma /d}Y$     & $\mathrm{d\sigma /d}Y$ & $\mathrm{d\sigma /d}Y$  \\ \hline
      $2.0-2.5$    &  $0.50 \pm 0.07 \pm 0.03$  & 0.40                   & 0.32 \\ \hline
      $2.5-3.5$    &  $0.30 \pm 0.03 \pm 0.02$  & 0.29                   & 0.17 \\ \hline
      $3.5-5.0$    &  $0.15 \pm 0.02 \pm 0.01$  & 0.14                   & 0.05 \\ \hline
      $5.0-7.0$    &  $0.08 \pm 0.02 \pm 0.01$  & 0.03                   & 0.006 \\ 
   \hline
  \end{tabular}}
\end{table}

\noindent

\begin{figure}[htbp]
  \begin{center}
  \includegraphics[height=0.25\textheight]{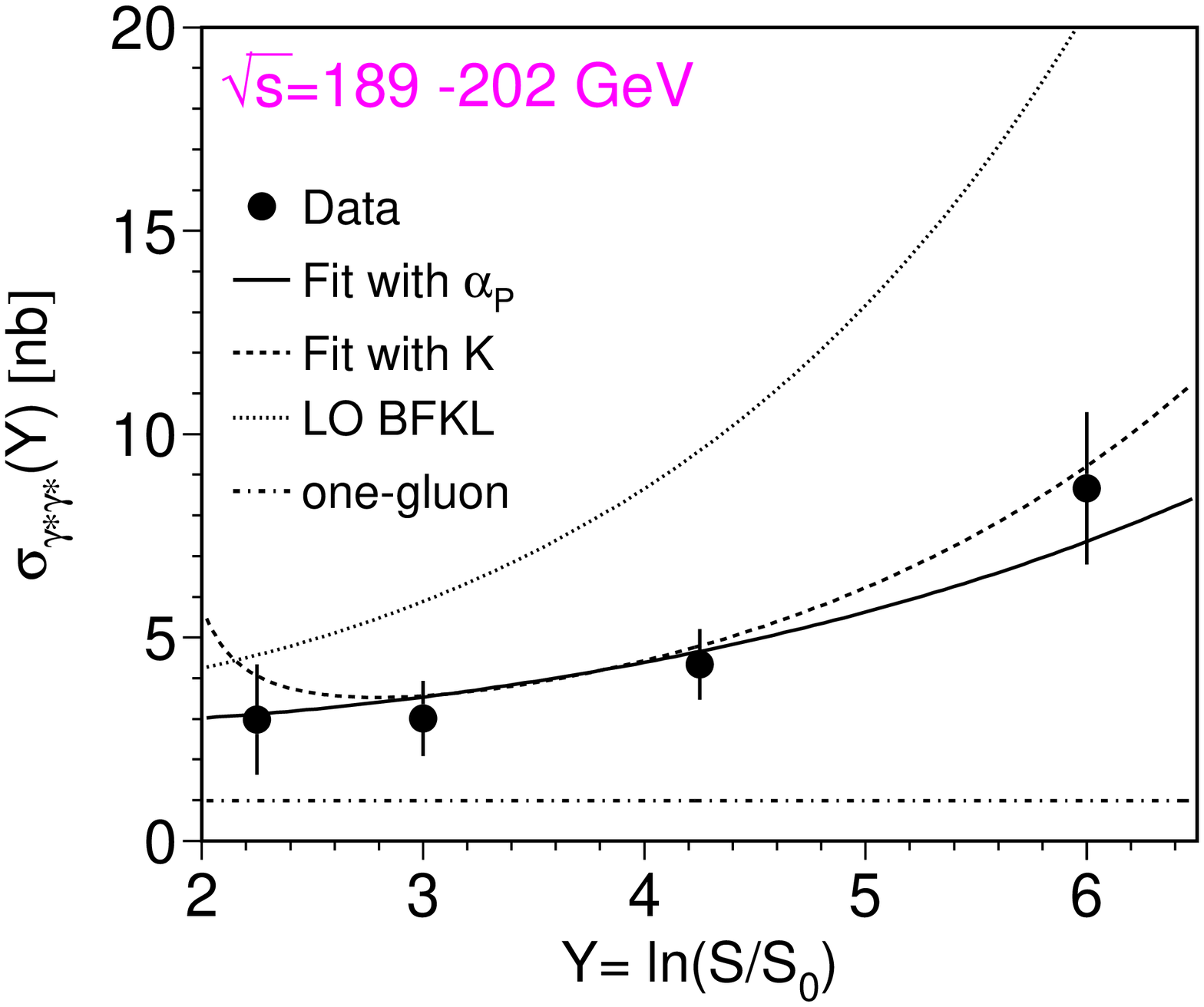}
  \caption[]{Two-photon cross-sections, $\sigma_{\gamma^*\gamma^*}$, after the subtraction of the
              QPM contribution at $\sqrt{s} \simeq 189-202 \GeV$~($\langle Q^2 \rangle = 15 \GeV^2$).
             The data are compared to the predictions of the LO BFKL calculation at saddle point
             approximation(eq.1)(dotted line) with K$=$1 and $(\alpha_P-1)=0.53$ and the solid line is the fit 
	     to the data of the LO BFKL (eq.1) with K$=$1 and the coefficient $(\alpha_P-1)$ as a 
	     free parameter. The dashed line is the fit with $(\alpha_P-1)=0.53$ and the scale factor K 
	     as a free parameter.
  \label{fig:ggcro} }
\end{center}
\end{figure}

From the measurement of the $\epem \rightarrow \epem + \mathrm{hadrons}$ cross-section, 
$\sigma_{\mathrm{ee}}$, we extract the two-photon cross-section, $\sigma_{\gamma^*\gamma^*}$, 
by using only the transverse photon luminosity function, $\sigma_{\mathrm{ee}} = L_{TT} \cdot  \sigma_{\GG}$.
In Fig.~\ref{fig:ggcro} we show $\sigma_{\GG}$, after subtraction of the QPM contribution
as a function of $Y$. Using an average value of $Q^2$, $\langle Q^2 \rangle =15 \GeV^2$ at
$\sqrt{s} \simeq 189 - 202\GeV$,   we calculate the one-gluon exchange contribution with the
asymptotic formula. The expectations are below the data. The leading order expectations of
the BFKL model,, shown as a dotted line in Fig.~\ref{fig:ggcro}, are too high. By
leaving $\alpha_P$ as a free parameter and $K=1$, a fit to the data, taking into account the
statistical, yields:

 $$\alpha_P - 1 = 0.36 \pm 0.02, ~~\chi^2/d.o.f =0.98/3$$

with $\chi^2$/d.o.f =0.98/3 and if the energy scale factor K is a free parameter 
and $(\alpha_P-1)$=0.53, a fit to data yields:

 $$K = 6.4 \pm 1.0, ~~\chi^2/d.o.f =1.34/3$$

These results are shown in Fig.~\ref{fig:ggcro} as a soild and dashed lines respectively.
The value of $(\alpha_P - 1)$, smaller than expected from the LO BFKL calculation at the saddle
point approximation, and the scale factor K much larger than unity indicate that higher order
corrections are important. NLO calculations are in progress~\cite{nlo7,nlo2,nlo6} which agree better 
with the experimental results.

\section*{Acknowledgements}
I would like to thank C. H. Lin of his collaboration. This work is supported by the 
Swiss National Science Foundation.

\end{document}
